\documentclass[aps,prd,final,twocolumn,letterpaper]{revtex4}

\usepackage{ez}

\usepackage{datetime}
\newdate{ogdate}{12}{05}{2023}

\begin{document}

\title{Analytic Solutions of Control Mechanism in Single-Qubit Systems}
\author{Erez Abrams}
\affiliation{Massachusetts Institute of Technology}
\date{\displaydate{ogdate}} 

\begin{abstract}
\noindent	
The mechanism governing the evolution of controlled quantum systems is often obscured, making their dynamics hard to interpret. Mitra and Rabitz {[Phys. Rev. A 67, 033407 (2003)]} define mechanism via a perturbative expansion of pathways between eigenstates; the evolution of the system is driven by the constructive and destructive interference of these pathway amplitudes. In this paper, we explore mechanism in controlled single-qubit systems and describe novel analytic methods for computing the mechanism underlying the evolution of a single qubit under a piecewise constant control.
\end{abstract}

\maketitle

\pagestyle{myheadings}
\markboth{E.~Abrams}{Analytic Solutions of Control Mechanism in Single-Qubit Systems}
\thispagestyle{empty}

\section{Introduction}\label{sec:intro}

Quantum control studies the use of external influences, such as shaped laser pulses or changing magnetic fields, in manipulating the dynamics of quantum systems. Quantum control has many applications, including ultracold physics, quantum computing, controlling chemical reactions, and nuclear magnetic resonance (NMR). Typically, control fields are designed either for population transfer (a specific transition is targeted so that if the system starts in a specific initial state $\ket i$, it will end in a desired final state $\ket f$) or for gate creation (an entire unitary time-evolution operator $U$ is targeted so that for \textit{all} initial states $\ket i$, the system will end the desired $U \ket i$). 

While in some cases, analytic construction of control fields is possible, in practice the process of designing controls is often opaque; the need to jointly optimize fidelity, fluence, and robustness makes gradient methods a reliable choice, but gradient methods rarely give insight into the means of their effectiveness. To remedy the lack of quantitative mechanistic information, Mitra and Rabitz \cite{abhra_1} described mechanism by decomposing the Dyson series expansion of the time-evolution operator into individual complex pathway amplitudes, each of which describes the mechanistic contribution of a unique pathway between eigenstates of the system. While these pathways may in principle be computed by numerically integrating terms of the Dyson series, this is usually prohibitively computationally expensive, so Mitra and Rabitz provide a toolbox of computational methods for extracting these pathways by modulating the Hamiltonian and decoding the evolution of these modulated systems \cite{abhra_1}.

In some special cases, the direct computation of mechanism can be simplified dramatically. One such case is when the Hamiltonian is constant in time, where the mechanism reduces to the Taylor expansion of a matrix exponential. This case becomes more useful when we observe that if we compute the control mechanism of two different pulses, we can compute the control mechanism of a concatenation of the two pulses by concatenating pathways from each pulse and multiplying their amplitudes. In this paper, we use this observation to provide novel analytic methods for computing the mechanism underlying the evolution of a single NMR qubit under a piecewise constant control.

\section{Mechanism Analysis}\label{sec:mechanism}
\subsection{Introduction to Mechanism}\label{ssec:mech-intro}

Quantum control studies the evolution of systems whose Hamiltonian $H = H_0 + H_I(t)$ is the sum of a field-free Hamiltonian $H_0$ and an interacting Hamiltonian $H_I(t)$ with zero diagonal which depends on the control field. The eigenstates of $H_0$ are denoted $\ket i$ with $H_0 \ket i = E_i \ket i$ for $i \in \qty{1, \dots, d}$, where $d$ is the dimension of the system.  In the interaction picture, the time-dependent Schr\"odinger equation is given as:
\begin{equation}
    i\hbar \dv{t}U(t) = V(t) U(t) \label{schro_eq}
\end{equation}
where $V(t) \equiv  \exp(iH_0t/\hbar)H_I(t) \exp(-iH_0t/\hbar)$. In the general case where $V(t)$ don't commute at different times, this equation is solved by the time-ordered exponential $U(t) = \mathcal T \exp (-\frac i \hbar \int_0^t V(t) \dd{t})$. Expanding this exponential yields the perturbative expansion for $U(t)$, known as the Dyson series, which is given as follows:
\begin{align*}\label{dyson_full}
     U&(t) = 1 + \qty(\frac{-i}{\hbar})\int_{0}^{t} V(t_1) \dd{t_1}\\  \notag
     &+ \qty(\frac{-i}{\hbar})^2\int_{0}^{t}\! \int_{0}^{t_2} V(t_2)  V(t_1) \dd{t_1} \dd{t_2}\\ \notag
     &+ \qty(\frac{-i}{\hbar})^3\int_{0}^{t}\! \int_{0}^{t_3}\! \int_{0}^{t_2} V(t_3)  V(t_2)  V(t_1) \dd{t_1} \dd{t_2}\dd{t_3}\\ \notag
     &+\ \cdots \nmberthis
\end{align*}
Noting that $\sum_{i=1}^d \dyad{i} = 1$, we insert a complete set of basis states in between every matrix product. Adopting the notation $U_{ba} = \mel{b}{U(T)}{a}$ and $v_{mn}(t) = - \frac i \hbar \mel{m}{V(t)}{n}= - \frac i \hbar \mel{j}{H_I(t)}{i} e^{i (E_j - E_i) t / \hbar}$ yields:
\begin{align*} \label{dyson_element}
    &U_{ba} = \delta_{ab} + \int_{0}^{T} v_{ba}(t_1) \dd{t_1}\\ \notag
     &+ \sum_{l=1}^d \int_{0}^{T}\!\!\! \int_{0}^{t_2} v_{bl}(t_2)  v_{la}(t_1) \dd{t_1} \dd{t_2}\\ \notag
     &+ \sum_{m=1}^d\! \sum_{n=1}^d \int_{0}^{T}\!\!\! \int_{0}^{t_3}\!\!\!  \int_{0}^{t_2}\!\! v_{bm}(t_3)  v_{mn}(t_2) v_{na}(t_1) \dd{t_1} \dd{t_2} \dd{t_3}\\ \notag
     &+ \cdots 
\end{align*}
We can define mechanism in terms of pathways directly from this expansion. We can express the transition amplitude $U_{ba}$ as:
\begin{equation*} \label{dyson_elements}
    U_{ba} = \sum_{n=0}^\infty \sum_{l_{n-1}=1}^d\cdots\sum_{l_1=1}^d U_{ba}^{n(l_1,\dots,l_{n-1})}
\end{equation*}
where $U^{0()}_{ba} = \delta_{ba}$ and:
\begin{align} \label{pathway_amps}
        U^{n(l_1,\dots,l_{n-1})}_{ba}\hspace{30pt}& \nonumber \\ 
        \equiv \int_0^T\int_0^{t_n}\cdots\int_0^{t_2} &v_{bl_{n-1}}(t_n)v_{l_{n-1}l_{n-2}}(t_{n-1})\cdots \nonumber \\ 
        & v_{l_1a}(t_1) \dd{t_1}\cdots\dd{t_{n-1}}\dd{t_n}
\end{align}
Each $U^{n(l_1,\dots,l_{n-1})}_{ba}$ is called a pathway amplitude for reasons which will soon become clear. The superscript $n$ denotes the \textit{order} of the amplitude, while the $l_i$ are intermediate states inserted in the expansion. It should be noted that no two adjacent states in a pathway should be equal because $V(t)$ has zero diagonal. Because each pathway amplitude is defined by a unique initial state, sequence of intermediate states, and ending state, we say that $U^{n(l_1,\dots,l_{n-1})}_{ba}$ is the amplitude of the $n$th-order perturbative pathway $\ket a \to \ket {l_1} \to \dots\to\ket{l_{n-1}}\to\ket b$. Pathways with large or small amplitudes have large and small contributions to the control dynamics, respectively, and different pathways constructively and destructively interfere to drive the control dynamics of the system.

\subsection{Single-Qubit Mechanism}\label{ssec:1q}
Before we even consider the Hamiltonian, the simplicity of the two-level single-qubit system means we should first inspect the possible pathways in the system. Notably, for each order, there is at most one valid pathway from a given start state to a given end state. For example, from $\ket 0$ to $\ket 1$, there is one first-order pathway $\ket0\to\ket1$, no second-order pathway, one third-order pathway $\ket0\to\ket1\to\ket0\to\ket1$, no fourth-order pathway, one fifth-order pathway $\ket0\to\ket1\to\ket0\to\ket1\to\ket0\to\ket1$, and so on. $\ket 0$ to $\ket0$ and $\ket1$ to $\ket1$ have only even pathways, while $\ket 0$ to $\ket1$ and $\ket1$ to $\ket0$ have only odd pathways, so we call the former ``even'' transitions and the latter ``odd'' transitions. 

We now turn to the Hamiltonian. A single qubit can be described by $H_0 = \omega_0S_z$ and $H_I = \epsilon_x(t) S_x + \epsilon_y(t) S_y$ so:
\begin{align*}
H(t) &= \omega_0S_z + \epsilon_x(t) S_x + \epsilon_y(t) S_y\\
V(t) &= e^{i \omega_0 S_z t /\hbar}(\epsilon_x(t) S_x + \epsilon_y(t) S_y) e^{-i \omega_0 S_z t /\hbar}
\end{align*}
We can expand this by noting that the exponentials act as rotation operators on the spin matrices:
\begin{align*}
V(t)&=\phantom{}(\epsilon_x(t) \cos\frac{\omega_0 t}{\hbar}+\epsilon_y(t) \sin\frac{\omega_0 t}{\hbar})S_x\\&\phantom{=}+(\epsilon_y(t) \cos\frac{\omega_0 t}{\hbar}-\epsilon_x(t) \sin\frac{\omega_0 t}{\hbar})S_y
\end{align*}
We can now define interaction fields $\tilde \epsilon_{x}(t), \tilde \epsilon_y(t)$ to obtain a much more convenient Hamiltonian:
\begin{align*}
    \tilde \epsilon_x(t) &= \epsilon_x(t) \cos\frac{\omega_0 t}{\hbar}+\epsilon_y(t) \sin\frac{\omega_0 t}{\hbar}\\
    \tilde \epsilon_y(t) &= \epsilon_y(t) \cos\frac{\omega_0 t}{\hbar}-\epsilon_x(t) \sin\frac{\omega_0 t}{\hbar}\\
    V(t) &= \tilde \epsilon_x(t) S_x + \tilde \epsilon_y(t) S_y
\end{align*}
Other than $\omega_0=0$ and a time-dependent rotation of the fields, this is identical to our original Hamiltonian. We can thus forget about $\omega_0$ entirely and work exclusively in this rotating interaction frame; we will do so moving forward, setting $\omega_0=0$, and as a result we will drop the tilde on $\tilde \epsilon_{x}= \epsilon_x, \tilde \epsilon_y=\epsilon_y$ and use $V$ synonymously with $H$.

With the field-free Hamiltonian eliminated and the powers of $S_x,S_y$ well-known, we can now expand the mechanism of a single pulse which is constant in time and active over a finite time interval. With interval $0\leq t < T$, we'll write the Hamiltonian as $H = \omega(S_x \cos\phi +  S_y\sin\phi) \equiv \omega S_\phi$ in the interval and $H=0$ otherwise, so that in the interval $H \ket 0 = \frac\hbar2\omega e^{i\phi}\ket 1$ and $H \ket 1 = \frac\hbar2\omega e^{-i\phi}\ket 0$. If we consider the mechanism from $t=0$ from $t=T$, we can forget about the transience of the pulse entirely and consider only the nonzero Hamiltonian. For the purpose of the expansion, we'll define $S_{\phi} = \frac\hbar2\sigma_\phi$ and note that $\sigma_\phi^2 =1$, so from Eq.~\ref{dyson_full} the Dyson series collapses to:
\begin{align*}
     U(T) &=1 + \qty(\frac{-i}{\hbar}) T\omega \frac\hbar2\sigma_\phi+\qty(\frac{-i}{\hbar})^2 \frac1{2!}T^2\omega^2 \frac{\hbar^2}{4}\\
     &+ \qty(\frac{-i}{\hbar})^3 \frac1{3!}T^3\omega^3 \frac{\hbar^3}{8}\sigma_\phi+\qty(\frac{-i}{\hbar})^4 \frac1{4!} T^4 \omega^4 \frac{\hbar^4}{16}+\ \cdots
\end{align*}
Since there's (at most) one pathway per order, we can read the pathways directly from this expansion. Since we need to distinguish even and odd order $n$, we'll introduce some new notation: $n_e$ and $n_o$ will indicate that $n$ is even or odd, respectively. Also, since $T$ and $\omega$ always show up together, we'll define the pulse width $\tau = T\omega$. Finally, since there's at most one pathway per order, we don't need the parentheses in the amplitude superscript. Directly from the expansion above, we conclude that the pulse with width $\tau$ and phase $\phi$ has mechanism:
\begin{align}\label{1q-mech-soln}
\notag U_{00}^{n_e} (T) &=\frac1{n!}\pqty{\frac{-i}2}^n \tau^n,& U_{00}^{n_o} (T) = 0\\
\notag U^{n_o}_{10}(T) &= \frac1{n!}\pqty{\frac{-i}2}^ne^{i\phi}\tau^n, & U^{n_e}_{10}(T)=0\\U^{n_e}_{11}(T) &= \frac1{n!}\pqty{\frac{-i}2}^n \tau^n
\notag,& U^{n_o}_{11}(T) = 0\\
U^{n_o}_{01}(T) &= \frac1{n!}\pqty{\frac{-i}2}^ne^{-i\phi}\tau^n, & U^{n_e}_{01}(T)=0
\end{align}
We will use this expansion extensively when we turn to piecewise constant controls. It also allows us to extrapolate that for any pulse sequence, $U_{00}^n = \overline{U_{11}^n}$ and $U_{10}^n = -\overline{U_{01}^n}$; we will not prove this here, but it is straightforwardly shown from the concatenation expansion of Section \ref{ssec:concat}.
\subsection{Mechanism Concatenation}\label{ssec:concat}
To expand our focus to piecewise constant pulses, we need to explore mechanism concatenation. Let's assume that we have two pulses whose lengths sum to $T$, $H(0<t<t_s) = H_1(t)$ and $H(t_s<t<T) = H_2(t)$. If $H_1, H_2$ are each zero outside their time interval, we can write this Hamiltonian as $H = H_1 + H_2$. Now we can expand the time-ordered exponential $U(T) = \mathcal T\exp\pqty{\frac{-i}{\hbar}\int_0^T(H_1(t) + H_2(t)) \dd {t}}$:
\begin{align*}
         U(T) = 1 &+ \qty(\frac{-i}{\hbar})\int_{0}^{T} (H_1(t_1)+H_2(t_1)) \dd{t_1}\\  \notag
     &+ \qty(\frac{-i}{\hbar})^2\int_{0}^{T}\! \int_{0}^{t_2} (H_1(t_2)+H_2(t_2))\\[-10pt] \notag
     &\phantom{+ \qty(\frac{-i}{\hbar})^2\int_{0}^{T}\! \int_{0}^{t_2} }\!\cdot(H_1(t_1)+H_2(t_1)) \dd{t_1} \dd{t_2}\\[-5pt] \notag
     &+\ \cdots
\end{align*}
Expanding each piece out by splitting the integrals on $t_s$, we see:
\begin{align*}
    \int_0^T &(H_1(t_1)+H_2(t_1))\dd{t_1} \\&= \int_0^{t_s} H_1(t_1)\dd{t_1} + \int_{t_s}^T H_2(t_1)\dd{t_1}
\end{align*}
and:
\begin{align*}
    \int_{0}^{t}\! \int_{0}^{t_2}&(H_1(t_2)+H_2(t_2) ) (H_1(t_1)+H_2(t_1) )\dd{t_1} \dd{t_2}  \\=& \int_{0}^{t_s}\!\int_{0}^{t_2} H_1(t_2)  H_1(t_1) \dd{t_1} \dd{t_2} \\&+ \int_{t_s}^t  H_2(t_2)\dd{t_2}\int_{0}^{t_s}  H_1(t_1) \dd{t_1} \\&+\int_{t_s}^t \int_{t_s}^{t_2} H_2(t_2)  H_2(t_1) \dd{t_1} \dd{t_2}\\
\end{align*}
and so on. We can prove the validity of this expansion in general by noting that $H_1$ and $H_2$ commute at any given time and splitting the time-ordered exponential. The validity of this expansion reflects the fact that we can write each order of the concatenated Dyson series by expanding a product of the Dyson series of the two individual pulses. This allows us to write the mechanism of the concatenated pulse as a convolution of the mechanisms of the individual pulses as well: for example, if $^1U$ and $^2U$ denote pathway amplitudes from the first and second pulses respectively, the pathway amplitude $U^{3(23)}_{41}$ of $\ket1\to\ket2\to\ket3\to\ket4$ in a four-level system would be:
\begin{align*}
    U^{3(23)}_{41} = {^1}U^{3(23)}_{41} + {^1}U^{2(2)}_{31}\,  {^2}U^{1()}_{43} +  {^1}U^{1()}_{21} \, {^2}U^{2(3)}_{42} + {^2}U^{3(23)}_{41}
\end{align*}
By repeating this concatenation process, we can describe the concatenation of any number of pulses. While pulse concatenation works in general, we generally calculate mechanism by encoding the Hamltonian rather than by direct computation \cite{abhra_1}, so this process is usually not any more efficient than simply encoding the Hamiltonian for the concatenated system. We explore piecewise constant single-qubit controls in this paper because we can analytically solve each constant pulse, allowing us to write down and manipulate closed-form expressions for the concatenated amplitudes.
\section{Piecewise Constant Single-Qubit Controls}\label{sec:pconst}
We can now finally turn to general piecewise constant controls; a pulse will henceforth refer to a constant control active over a finite time interval. In this section, we will detail a general procedure for analytically calculating pathway implitudes on an arbitrary piecewise-constant single-qubit control. We'll start with two pulses, then three pulses, and then we'll describe a general procedure for $M$ pulses. As the mechanism depends only on the pulse phase and width, we'll define each pulse $k$ by its width $\tau_k$ and phase $\phi_k$. We will also now let $N$ be the order to free up $n$ for use as a summation index. Since  $U_{11}^N = \overline{U_{00}^N}$ and $U_{01}^N = -\overline{U_{10}^N}$, we can just expand $U_{00}^{N_e}$ and $U_{10}^{N_o}$. 
\subsection{Two Pulses}\label{ssec:2p}
The two-pulse case is a special case because it lends itself to a neat final result which unfortunately does not generalize to higher numbers of pulses. We'll start with $U_{10}^{N_o}$, noting the odd and even subscripts in the summations. Applying the techniques of Section \ref{ssec:concat} to Eq.~\ref{1q-mech-soln}:
\begin{align*}
    U&^{N_o}_{10}(T) \\&= \qty(\frac{-i}2)^N\left( \sum_{n_o<N} \frac1{n!} e^{i\phi_1} \tau_1^n \frac1{(N-n)!}  \tau_2^{N-n}\right. \\&\qquad\qquad\quad\left.+ \sum_{n_e<N} \frac1{n!} \tau_1^n \frac1{(N-n)!} e^{i\phi_2} \tau_2^{N-n} \right)
    \\&= \qty(\frac{-i}2)^N\left(\frac1{N!} \sum_{n_o}\binom{N}{n}\tau_1^n   \tau_2^{N-n}e^{i\phi_1} \right. \\&\qquad\qquad\quad\left.+ \frac1{N!}\sum_{n_e} \binom{N}{n} \tau_1^n \tau_2^{N-n} e^{i\phi_2}  \right)\\
\end{align*}
We can deal with the odd and even sums by defining $\tau_+ = \tau_1+\tau_2$ and $\tau_- = -\tau_1 + \tau_2$ so that:
\begin{align*}\label{eq:s0s1}
\tau_+^N&=(\tau_1+\tau_2)^N=  \sum_{n} \binom{N}{n} \tau_1^n   \tau_2^{N-n}\\
 \tau_-^N&= (-\tau_1 + \tau_2)^N =  \sum_{n} \binom{N}{n} (-\tau_1)^n   \tau_2^{N-n} \nmberthis
\end{align*}
These polynomials allow us to concisely describe the odd and even sums:
\begin{align*}
    \sum_{n_e<N} \frac1{n!} \tau_1^n \frac1{(N-n)!}  \tau_2^{N-n}&=\frac1{N!}\sum_{n_e}\binom{N}{n}\tau_1^n   \tau_2^{N-n}\\ &= \frac12 \frac1{N!}(\tau_+^N + \tau_-^N)\\
        \sum_{n_o<N} \frac1{n!} \tau_1^n \frac1{(N-n)!}  \tau_2^{N-n}&=\frac1{N!}\sum_{n_o}\binom{N}{n}\tau_1^n   \tau_2^{N-n}\\ &= \frac12 \frac1{N!}(\tau_+^N - \tau_-^N)
\end{align*}
We can now proceed with the pathway amplitude:
\begin{align*}
        U&_{10}^{N_o}(T) \\&=\!\pqty{\frac{-i}2}^N\!\!\pqty{e^{i\phi_1} \frac1{2}\frac1{N!}(\tau_+^N-\tau_-^N) + \!e^{i\phi_2} \frac1{2}\frac1{N!}(\tau_+^N+\tau_-^N) \!}\\
        &= \!\pqty{\frac{-i}2}^N\!\frac1{N!}\frac12\pqty{e^{i\phi_1} \tau_+^N  - e^{i\phi_1} \tau_-^N + e^{i\phi_2} \tau_+^N + e^{i\phi_2} \tau_-^N }\\
         &=\!\pqty{\frac{-i}2}^N\!\frac1{N!}\frac12 ((e^{i\phi_1}+e^{i\phi_2})(\tau_1 + \tau_2)^N \\&\qquad\qquad\qquad\quad+ (-e^{i\phi_1 } + e^{i\phi_2} ) (-\tau_1 + \tau_2)^N)\\
\end{align*}
This form allows us to concisely evaluate the pathway amplitude without performing sums over $O(N)$ terms. For larger $M$, this will be our final form, but we can proceed to a more pleasing expression in the two-pulse case by defining $\phi_+ = (\phi_1+\phi_2)/2$ and $\phi_-=(\phi_1-\phi_2)/2$:
\begin{align*}
    U_{10}^{N_o}(T) = \!\pqty{\frac{-i}2}^N\!\frac1{N!} e^{i\phi_+}( &(\tau_1+\tau_2)^N\cos\phi_- \\+ &(-\tau_1+\tau_2)^Ni\sin\phi_-)
\end{align*}

A similar expansion for $U_{00}^{N_e}(T)$ gives:
\begin{align*}
    U&^{N_e}_{00}(T) \\&= \qty(\frac{-i}2)^N\left( \sum_{n_o<N} \frac1{n!} e^{i\phi_1} \tau_1^n \frac1{(N-n)!}  e^{-i\phi_2}\tau_2^{N-n}\right. \\&\qquad\qquad\quad\left.+ \sum_{n_e<N} \frac1{n!} \tau_1^n \frac1{(N-n)!}  \tau_2^{N-n} \right)\\
         &=\!\pqty{\frac{-i}2}^N\!\frac1{N!}\frac12 ((1+e^{i(\phi_1-\phi_2)})(\tau_1 + \tau_2)^N \\&\qquad\qquad\qquad\quad+ (1 - e^{i(\phi_1-\phi_2)} ) (\tau_1 - \tau_2)^N)\\   
         &= \!\pqty{\frac{-i}2}^N\! \frac1{N!} e^{i\phi_-} ( (\tau_1+\tau_2)^N\cos\phi_- \\
         &\qquad\qquad\qquad\quad+(-\tau_1+\tau_2)^Ni\sin\phi_-)
\end{align*}
These are the analytic solutions for the mechanism of the two-pulse case. Taking $\tau_2=0$ or $\phi_1=\phi_2$ reduces this to the one-pulse case and recovers Eq.~\ref{1q-mech-soln} as expected.
\subsection{Three Pulses}
We will now proceed with the three-pulse case in largely the same way as the two-pulse case. Some choices of notation may seem odd at first, but they are chosen to be suggestive of the general case. We start by applying Section \ref{ssec:concat} to Eq.~\ref{1q-mech-soln} twice:
\begin{align*}\label{eq:3p-sigma}
    U^{N_o}_{10}(T)\qquad& \\= \pqty{\frac{-i}2}^N&\left( \sum_{n_e+m_e+l_o=N} \frac1{n!m!l!} \tau_1^n\tau_2^m\tau_3^l e^{i\phi_3}\right.
\\+&\sum_{n_o+m_e+l_e=N} \frac1{n!m!l!} \tau_1^n\tau_2^m\tau_3^l e^{i\phi_1}\\+&\sum_{n_e+m_o+l_e=N} \frac1{n!m!l!} \tau_1^n\tau_2^m\tau_3^l e^{i\phi_2}\\+&\left.\sum_{n_o+m_o+l_o=N} \frac1{n!m!l!} \tau_1^n\tau_2^m\tau_3^l e^{i(\phi_1-\phi_2+\phi_3)}\right)\\
\equiv \!\pqty{\frac{-i}2}^N &\!\frac1{N!}(\Sigma_\qty{} e^{i\phi_3} \!+\!\Sigma_\qty{1} e^{i\phi_1} \\&\quad+\Sigma_\qty{2} e^{i\phi_2} \!+\!\Sigma_\qty{1,2} e^{i(\phi_1\!-\!\phi_2\!+\!\phi_3)} )\nmberthis\\
\end{align*}
Where the sets in the $\Sigma$ subscripts describe which of the first two integers in the sum are odd. We once again deal with the odd and even sums by defining polynomials in $\tau$. Using the multinomial coefficients $\binom{N}{n,m,l} = \frac{N!}{n!m!l!}$ with $n+m+l=N$, we write:
\begin{align*}\label{eq:tauset}
    \tau_\Bqty{}^N &\equiv (\tau_1+\tau_2+\tau_3)^N \\&= \sum_{n+m+l=N} \binom{N}{n,m,l} \tau_1^n\tau_2^m\tau_3^l\\
    \tau_\Bqty{1}^N &\equiv (-\tau_1+\tau_2+\tau_3)^N \\&= \sum_{n+m+l=N} \binom{N}{n,m,l} (-\tau_1)^n\tau_2^m\tau_3^l\\
    \tau_\Bqty{2}^N &\equiv (\tau_1-\tau_2+\tau_3)^N \\&= \sum_{n+m+l=N} \binom{N}{n,m,l} \tau_1^n(-\tau_2)^m\tau_3^l\\
    \tau_\Bqty{1,2}^N& \equiv (-\tau_1-\tau_2+\tau_3)^N \\&= \sum_{n+m+l=N} \binom{N}{n,m,l} (-\tau_1)^n(-\tau_2)^m\tau_3^l\nmberthis
\end{align*} 
Where the sets in the $\tau$ subscripts describe which $\tau_i$ are negated. Such powers of $\tau$ sums can be written as a linear combination of $\Sigma$ sums as in Eq.~\ref{eq:s0s1}:
\begin{align*}\label{eq:3p-tau}
\tau_{\qty{}}^{N_o}=(\tau_1+\tau_2+\tau_3)^{N_o} &=\Sigma_\qty{} + \Sigma_\qty{1} + \Sigma_\qty{2} + \Sigma_\qty{1,2} \\
\tau_{\qty{1}}^{N_o}=(-\tau_1+\tau_2+\tau_3)^{N_o} &=\Sigma_\qty{} - \Sigma_\qty{1} + \Sigma_\qty{2} - \Sigma_\qty{1,2}\\
\tau_{\qty{2}}^{N_o}=(\tau_1-\tau_2+\tau_3)^{N_o} &=\Sigma_\qty{} + \Sigma_\qty{1} - \Sigma_\qty{2} - \Sigma_\qty{1,2}\\
\tau_{\qty{1,2}}^{N_o}=(-\tau_1-\tau_2+\tau_3)^{N_o} &=\Sigma_\qty{} - \Sigma_\qty{1} - \Sigma_\qty{2} + \Sigma_\qty{1,2}\nmberthis\\
\end{align*}
This system is inverted as:
\begin{align*}\label{eq:3p-invert}
4\Sigma_{\qty{}} &=\tau^{N_o}_\qty{} + \tau^{N_o}_\qty{1} + \tau^{N_o}_\qty{2} + \tau^{N_o}_\qty{1,2} \\
4\Sigma_{\qty{1}}&=\tau^{N_o}_\qty{} - \tau^{N_o}_\qty{1} + \tau^{N_o}_\qty{2} - \tau^{N_o}_\qty{1,2}\\
4\Sigma_{\qty{2}}&=\tau^{N_o}_\qty{} + \tau^{N_o}_\qty{1} - \tau^{N_o}_\qty{2} - \tau^{N_o}_\qty{1,2}\\
4\Sigma_{\qty{1,2}} &=\tau^{N_o}_\qty{} - \tau^{N_o}_\qty{1} - \tau^{N_o}_\qty{2} + \tau^{N_o}_\qty{1,2}\nmberthis\\
\end{align*}
So we can now write:
\begin{align*}\label{eq:3p-final}
    U^{N_o}_{10}(T)\qquad&\\
     = \!\pqty{\frac{-i}2}^N \!\frac1{N!}\hspace{-1.5pt}&\hspace{1.5pt}\frac14(\tau^{N_o}_\qty{}{} (e^{i\phi_3}+e^{i\phi_1}+e^{i\phi_2}+e^{i(\phi_1-\phi_2+\phi_3)}) \\&+ \tau^{N_o}_\qty{1}{}(e^{i\phi_3}-e^{i\phi_1}+e^{i\phi_2}-e^{i(\phi_1-\phi_2+\phi_3)}) \\&+ \tau^{N_o}_\qty{2}{}(e^{i\phi_3}+e^{i\phi_1}-e^{i\phi_2}-e^{i(\phi_1-\phi_2+\phi_3)}) \\&+ \tau^{N_o}_\qty{1,2}{}(e^{i\phi_3}-e^{i\phi_1}-e^{i\phi_2}+e^{i(\phi_1-\phi_2+\phi_3)})) \nmberthis\\
\end{align*}
This reflects the general form that we will see in the next section. The case of $U_{00}^{N_e}(T)$ is solved similarly, so we will not detail it here.
\subsection{Arbitrary Pulse Sequneces}\label{ssec:arbit}
We will now describe a general procedure for calculating any pathway amplitude $U_{ba}^N$ for a piecewise-constant pulse on a single qubit. This procedure will assume that the parity of $N$ matches the parity of the transition (e.g.~$U_{10}^N$ should have $N$ odd) since the amplitude is zero otherwise. We'll break up the piecewise-constant pulse into $M$ constant pulses, where each pulse has width $\tau_m$ and phase $\phi_m$. To prepare, we will define the powerset $\mathcal P_{M-1}$ as the set of all subsets of $\Bqty{1,\dots,M-1}$; each $S \in \mathcal P_{M-1}$ is a set of integers $s_i$ with $1 \leq s_1< \dots< s_{|S|} \leq M-1$.
\begin{enumerate}

\item Mirroring Eq.~\ref{eq:3p-sigma}, for each $S\in \mathcal P_{M-1}$ we define a sum $\Sigma_S$ over sets over integers $\Bqty{n_1,\dots, n_M}$ satisfying $\sum_m  n_m = N$, where for $m<M$ we have $n_m$ is odd for $m \in S$ and even for $m \notin S$, and $n_M$ takes the necessary parity for $\sum_m n_m = N$ ($n_M$ is even if $|S| \equiv N\mod 2$, odd otherwise): 
\begin{equation}
\Sigma_S \equiv \sum_{\substack{\sum_m n_m=N, \\ \text{parity defined by $S$}}}\binom{N} { n_1,\dots,n_M}\prod _k\tau_k^{n_k}
\end{equation}

\item For each $S$ we define a phase $\phi_S$. This phase is defined slightly differently depending on the parities of $N$ and $|S|$:
   \begin{align}
   \phi_S \equiv \begin{cases}\displaystyle
   \sum_{0<k\leq|S|}(-1)^{k-1}\phi_{s_k} &\text{$|S|$ odd, $N$ odd}\\
   \displaystyle\sum_{0<k\leq|S|}(-1)^{k-1}\phi_{s_k}  + \phi_M&\text{$|S|$ even, $N$ odd}\\
   \displaystyle
   \sum_{0<k\leq|S|}(-1)^{k-1}\phi_{s_k} -\phi_M &\text{$|S|$ odd, $N$ even}\\
   \displaystyle\sum_{0<k\leq|S|}(-1)^{k-1}\phi_{s_k}  &\text{$|S|$ even, $N$ even}
   \end{cases}
   \end{align}
   
\item With both definitions above, we can now write the pathway amplitude in a form mirroring Eq.~\ref{eq:3p-sigma}:
\begin{equation}
U_{ba}^{N} = \pqty{\frac{-i}2}^N \frac1{N!} \sum_{S \in \mathcal P_{M-1}} \Sigma_S e^{i\phi_S}
\end{equation}

\item Let $I_S (k) = 1$ if $k \in S$ and $I_S(k)=0$ otherwise. Mirroring Eq.~\ref{eq:tauset}, we define a $\tau_S$ for each $S$:
\begin{equation}
    \tau_S \equiv \sum_{0<k\leq M} (-1)^{I_S(k)}\tau_k 
\end{equation}
\item Mirroring Eq.~\ref{eq:3p-tau}, each $\tau_S^N$ can be uniquely written as a linear combination of $\Sigma_S$:
\begin{equation}\label{eq:fwht}
   \tau_S^N = \sum_{S' \in \mathcal P_{M-1}} (-1)^{|S\cap S'|}\Sigma_{S'}
   \end{equation}

\item Under the natural order \footnote{This order can be derived by counting in binary, i.e.~by 
placing $S$ before $S'$ if $\sum_{0 <k \leq |S|} 2^{s_k} < \sum_{0 < k \leq |S'|} 2^{s_k}$.} $\Bqty{}$, $\Bqty{1}$, $\qty{2}$, $\qty{2,1}$, $\qty{3}$,~$\dots$, the matrix coefficients of Eq.~\ref{eq:fwht} form a Hadamard matrix, so we can invert this directly mirroring Eq.~\ref{eq:3p-invert}:
\begin{equation}
   \Sigma_S = \frac1{2^{M-1}}\sum_{S' \in \mathcal P_{M-1}} (-1)^{|S\cap S'|}\tau _{S'}^N
   \end{equation}

\item Now, following Eq.~\ref{eq:3p-final}, we can write:
   \begin{align*}\label{eq:finalarb}
   &U_{ba}^{N} = \pqty{\frac{-i}2}^N \frac1{N!} \sum_{S \in \mathcal P_{M-1}} \Sigma_S e^{i\phi_S}\\
   &=  \pqty{\frac{-i}2}^N \frac1{N!} \frac1{2^{M-1}}\sum_{S \in \mathcal P_{M-1}} \sum_{S' \in \mathcal P_{M-1}} (-1)^{|S\cap S'|}\tau _{S'} ^Ne^{i\phi_S}\\
    &= \pqty{\frac{-i}2}^N \frac1{N!} \frac1{2^{M-1}} \sum_{S \in \mathcal P_{M-1}} \!\tau_S^N\!\sum_{S' \in \mathcal P_{M-1}}(-1)^{|S\cap S'|} e^{i \phi_{S'}}\nmberthis
   \end{align*}
\end{enumerate}
This procedure gives a general solution for piecewise constant pulses on a one-qubit system. Since Eq.~\ref{eq:fwht} takes the form of a Hadamard matrix, Eq.~\ref{eq:finalarb} can be evaluated by a fast Walsh-Hadamard transform in $O(M2^M)$ time. This is significant both because it does not depend on $N$ and because it does not involve numerical methods, whereas directly evaluating the Dyson terms of an $N$th-order pathway usually requires computationally expensive $N$-dimensional numerical integration over time.

\section{Discussion}
We have now fully explored the mechanism of a single qubit driven by an arbitrary $M$-segment piecewise constant control. The procedure developed in Section \ref{ssec:arbit} yields an analytic expression for the amplitude of any pathway on this system that can be evaluated in $O(M 2^M)$ time, which is notably independent of the order of the calculated pathway amplitude. Analytically designed robust single-qubit gates are often built by concatenating a small number of constant pulses (\cite{composite} uses $3 \leq M \leq 9$); since evaluating the analytic expression takes time independent of pathway order, the procedure provided is useful and practical for understanding the mechanism driving the evolution of a qubit under these controls. Gradient methods of control development often use large $M$; due to its rapidly growing runtime with respect to $M$, this procedure is not practical to implement for such controls, but it is still much faster than direct computation of the Dyson series terms because it does not involve the usual high-dimensional numerical integration. Regardless of its practicality, this solution is pedagogically interesting as the mechanism for a system with nonconstant Hamiltonian has never been solved analytically before.

\subsection*{Acknowledgments}
 The author is grateful to Michael Kasprzak for his support during the development of the two-pulse and three-pulse solutions, to Tomasz \'Slusarczyk and Nina Anikeeva for discussing and reviewing this paper, and to Herschel Rabitz and Gaurav Bhole for presenting this mechanistic problem.

\bibliography{OHPEsources}

\begin{thebibliography}{2}
\expandafter\ifx\csname natexlab\endcsname\relax\def\natexlab#1{#1}\fi
\expandafter\ifx\csname bibnamefont\endcsname\relax
  \def\bibnamefont#1{#1}\fi
\expandafter\ifx\csname bibfnamefont\endcsname\relax
  \def\bibfnamefont#1{#1}\fi
\expandafter\ifx\csname citenamefont\endcsname\relax
  \def\citenamefont#1{#1}\fi
\expandafter\ifx\csname url\endcsname\relax
  \def\url#1{\texttt{#1}}\fi
\expandafter\ifx\csname urlprefix\endcsname\relax\def\urlprefix{URL }\fi
\providecommand{\bibinfo}[2]{#2}
\providecommand{\eprint}[2][]{\url{#2}}

\bibitem[{\citenamefont{Mitra and Rabitz}(2003)}]{abhra_1}
\bibinfo{author}{\bibfnamefont{A.}~\bibnamefont{Mitra}} \bibnamefont{and}
  \bibinfo{author}{\bibfnamefont{H.}~\bibnamefont{Rabitz}},
  \bibinfo{journal}{Phys. Rev. A} \textbf{\bibinfo{volume}{67}},
  \bibinfo{pages}{033407} (\bibinfo{year}{2003}),
  \urlprefix\url{https://link.aps.org/doi/10.1103/PhysRevA.67.033407}.

\bibitem[{\citenamefont{Jones}(2013)}]{composite}
\bibinfo{author}{\bibfnamefont{J.~A.} \bibnamefont{Jones}},
  \bibinfo{journal}{Phys. Rev. A} \textbf{\bibinfo{volume}{87}},
  \bibinfo{pages}{052317} (\bibinfo{year}{2013}),
  \urlprefix\url{https://link.aps.org/doi/10.1103/PhysRevA.87.052317}.

\end{thebibliography}

\end{document}